\documentclass[%
 reprint,
 amsmath,amssymb,
 aps,
]{revtex4-2}

\usepackage{graphicx}
\usepackage{dcolumn}
\usepackage{bm}

\usepackage[center]{subfigure}

\begin{document}

 \newcommand{\bq}{\begin{equation}}
 \newcommand{\eq}{\end{equation}}
 \newcommand{\bqn}{\begin{eqnarray}}
 \newcommand{\eqn}{\end{eqnarray}}
 \newcommand{\nb}{\nonumber}
 \newcommand{\lb}{\label}
\newcommand{\PR}{Phys. Rev. }

\title{Dust as probes: determining confinement and interaction forces}

\author{Khandaker Sharmin Ashrafi$^1$}

\author{Razieh Yousefi$^2$}

\author{Mudi Chen$^1$}

\author{Lorin S. Matthews$^1$}
\email{Lorin\_Matthews@baylor.edu}

\author{Truell W. Hyde$^1$}

\affiliation{$^1$CASPER, Baylor University, Waco, TX 76798-7316, USA\\
$^2$The University of Texas Medical Branch, Galveston, TX 77555, USA}
\date{\today}

\begin{abstract}
Complex plasmas are interesting systems as the charged dust can self-assemble into different types of ordered structures. To understand the mechanisms which govern the transitions from one type of structure to another, it is necessary to know both the dust charge and the confining electric fields within the environment, parameters which are difficult to measure independently. As dust is usually confined in a plasma sheath where the ions stream from the bulk plasma the negative lower electrode, the problem is further complicated by the ion wake field, which develops downstream of the dust grains in a flowing plasma. The differences in local ion density caused by the wake field change the equilibrium dust charge and shielding distance of the dust grains, and thus affect the interaction between grains. Here we use a molecular dynamics simulation of ion flow past dust grains to investigate the interaction between the dust particles and ions. We consider a long vertical chain of particles confined within a glass box placed on the lower electrode of a GEC rf reference cell.  We apply the model iteratively to self-consistently determine the dust charge, electric field, and ion density along the length of the chain as well as the ion flow speed. Simulation results indicate that the ion flow speed within the box is subsonic.  

\end{abstract}


\maketitle

\section*{I. INTRODUCTION}

Dusty (complex) plasmas have attracted increasing interest in the last several years due to their presence in numerous space and astrophysical environments as well as fusion devices and industrial processes \cite{1a,1b,1c,1d}. In industrial environments, low-temperature dusty plasmas are extensively used in the manufacturing of chips and material processing, as well as in the production of thin films and nanoparticles \cite{1cc,1ccc}.

Dust particles charge by collecting electrons and ions from the plasma, and generally become negatively charged due to the higher mobility of the electrons \citep{4}.  In laboratory rf plasmas, these particles are most often trapped in the sheath of the plasma, above the lower electrode where an inhomogeneous electric field in the vertical direction levitates the particles against gravity \citep{5,6,8}. Due to the strong confinement  provided by the plasma sheath in the vertical direction, ordered structures tend to be two-dimensional, forming either a plasma crystal \citep{9,10,10a,11,12,1f} or horizontal clusters \cite{12a}. Placing an open glass box on the lower electrode provides increased horizontal confinement, allowing the formation of dust structures which are extended in the vertical direction. The horizontal confinement can be adjusting by changing the system power and/or pressure, allowing single or multiple vertical chains to be formed \cite{12aa}.

The stability of vertically aligned structures is not easily explained on the basis of a repulsive interparticle potential alone and appears to be dependent on the balance between an attractive ion wakefield and repulsive screened-Coulomb potentials \citep{3,13}. In the region of the sheath-plasma interface above the lower electrode, positive ions are accelerated downward toward the lower electrode. The resulting ion flow, deflected by the negatively charged grains, leads to the formation of excess positive space charge just below the negatively charged dust grains levitated in the sheath \citep{14}. Previous numerical simulations have shown that the positive space charge becomes weaker for downstream grains, as a significant fraction of ions are deflected by upstream grains so that their direction of motion is no longer along the direction of unperturbed ion flow \cite{15}. In addition, the downstream particles have a smaller charge, resulting in a smaller electrostatic lens \citep{15a}.   

In order to explain the stability of the dust structures formed within a glass box, one must determine the charge on each grain, the resultant interaction between dust and ions, and the electric field as well as the ion flow speed inside the box, all of which are difficult to measure experimentally. This is further complicated by the fact that extended vertical dust structures can span a significant fraction of the sheath with varying plasma conditions. However, the detailed balance which is required for stable structures makes micron-sized dust particles very sensitive to small changes in plasma conditions \cite{12aa}.  In addition, the dust particles are minimally perturbative to the plasma environment \cite{12aaa}, allowing them to be used as probes to investigate both the confinement and interaction forces \citep{12aa,12n}.

In this study, we take advantage of the large vertical extent provided by a 1D seven-particle dust chain, stably formed at a given power and pressure setting, to probe the electric field and ion flow within the glass box. The experimentally measured particle positions and global plasma conditions are used as inputs in a molecular dynamics simulation which models the flowing ions, the formation of the ion wake below the particles, and the charging of the dust particles \citep{15bbb}. An iterative procedure is used to self-consistently determine the local plasma conditions defining the sheath electric field which accelerates the ions and supports the dust against gravity, the speed of the flowing ions, and the charge of each dust grain. This analysis, which examines a single static configuration of the dust, will provide a basis for further study of the dust-plasma interactions which cause transitions from one stable configuration to another. 

The paper is organized as follows. In Section II, a short description of the experimental setup, along with the forces acting on each dust grain, is briefly discussed.  The details of the numerical model, including the forces acting on the ion, dust grains, and the calculations of the dust charge is discussed in Section III. Section IV describes the iterative approach used to determine the sheath electric field and ion flow velocity, and the results obtained using three different sets of initial conditions. A discussion of the results is provided in Section V, with a conclusion and direction for future work presented in Section VI.

 \section*{II. EXPERIMENT} 

\begin{figure}
\includegraphics[width=8cm]{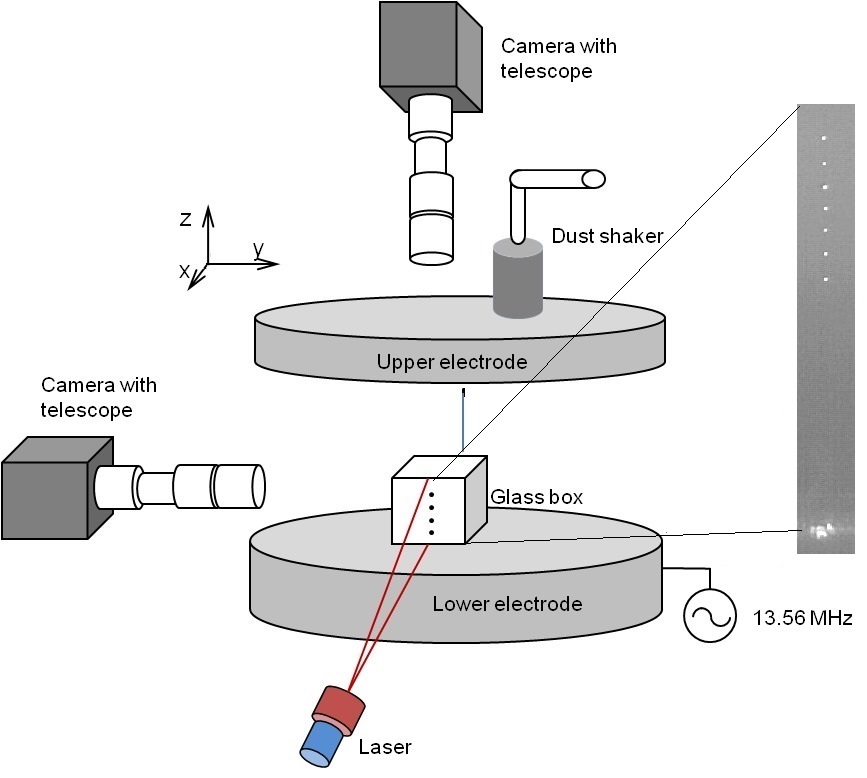}
\centering
\caption{(Color online) A schematic of the experimental setup. The open-ended glass box sitting on the lower electrode has dimensions of 12 mm $\times$ 10.5 mm $\times$ 10.5 mm. Dust particle chains of different lengths can be formed inside the box by adjusting the operating power at a given pressure. A vertical chain of seven particles providing the particle positions for the numerical analysis is shown at right. }
\label{fig1}
\end{figure}

The experimental data used here were obtained from an experiment conducted in a modified GEC (Gaseous Electronics Conference) rf reference cell. The cell has two electrodes: a lower cylindrical electrode of diameter 8 cm driven at 13.56 MHz and a hollow cylindrical upper electrode, which is grounded. The electrodes are separated from each other by a distance of 2.54 cm. As shown in Fig. \ref{fig1}, an open-ended glass box of dimension 12 mm $\times$ 10.5 mm $\times$ 10.5 mm (height $\times$ width $\times$ length) with 2 mm wall thickness is placed on top of the lower electrode in order to provide horizontal confinement for the dust particles \citep{11}. Melamine Formaldehyde (MF) spheres with a manufacturer provided diameter of 8.89 $\pm$ 0.09 $\mu$m and mass of 5.56 $\pm$ 0.09$ \times 10^{-13}$ kg were introduced into the argon plasma, using a shaker mounted above the hollow upper electrode, and imaged at 500 frames per second  using a side-mounted, high-speed CCD (Photron) camera and a microscope lens. Slowly lowering the rf power causes the dust cloud to shrink in the horizontal direction and stretch in the vertical direction, reducing the number of trapped particles as they are lost to the lower electrode.  The power was adjusted to form a single vertical dust chain consisting of seven dust grains spanning a vertical distance of 2.66 mm at the center of the glass box with the rf power set at 1.6 W at a gas pressure of 150 mTorr. Langmuir probe measurements in the bulk taken for similar conditions yielded a plasma density of $2\times 10^{15}$ $m^{-3}$ and an electron temperature $T_e \approx 5$ $eV$ \citep{15dc}, though the plasma density in the sheath will be reduced from this level.

The stable structure of the chain indicates that the total force acting on each particle is approximately zero. The total force on each particle assuming a charge $ Q_d $ and mass $m_d$ is 
\begin{equation}
\label{dust_force}
 m_d \vec g +  Q_d \vec E + \vec F_{dd} + \vec F_{di} = 0
\end{equation}
where $m_d \vec g $ is the gravitational force, $Q_d \vec E $ is the force from the sheath electric field in the confining region (and is the same electric field as that acting on the ions; See Eqns. \ref{sheath_Ez} and \ref{force_outside_ions}), $\vec F_{dd} $ is the force exerted by all other dust grains, $\vec F_{di} $ is the force from the ions.

The electric field within the sheath is generally assumed to be a linear function of the distance from the lower electrode \citep{15dc1}. However, within the glass box, the sheath is altered by the charge collected on the walls of the box \citep{15bb}. In this case, the sheath electric field is found to be a non-linear function of $z$ \citep{15bb}, and the ion velocity is not known with certainty. Thus, a numerical model is used here to calculate $\vec F_{di} $ and $\vec F_{dd} $ for given flow speeds, where the positions of the dust grains are those measured in the experiment. The electric field in the region spanned by the dust chain is then determined by solving Eq. (\ref{dust_force}) for each of the dust grains. This new electric field is then used in the simulation (and to estimate a new ion flow speed for ions entering the simulation region from above), and the procedure is iterated until the results converge. 

\section*{III. NUMERICAL MODEL}

We employ the molecular dynamics simulation DRIAD (Dynamic Response of Ions And Dust) to model the ion flow while simultaneously calculating the charge on the dust particles $Q_d$, the forces from the ions $ \vec  F_{di}$, and computing the effects of the ion wake downstream of each particle. A detailed description of the code is provided in \citep{15bbb}. A cylindrical simulation region surrounding the dust chain with the axis aligned with the ion flow is considered. The total number of simulated particles is reduced by using superions comprised of $N_i$ real ions that have the same charge to mass ratio, and hence dynamics, as a single ion. Each ion is initially given a random position and velocity inside the simulation cylinder. The vertical electric field within the plasma sheath above the lower electrode accelerates the ions downwards, towards the electrode.  Accordingly, ions enter the top and sides of the cylindrical region with a shifted Maxwellian velocity distribution \citep{15bbbb}. The interaction between ions is assumed to be a shielded Yukawa interaction, while the interaction between the ions and the dust particles is asymmetric (i.e. the force on the ions from the dust is derived assuming a Coulomb potential, while the force on the dust from the ions is derived assuming a Yukawa potential). This asymmetric interaction, when used in a MD simulation, allows the nonlinear shielding of the grain due to the ions to be addressed at the expense of an approximate treatment of electron shielding, and has been shown to yield good agreement with PIC simulations of the ion wake \citep{15b}.

The equation of motion for each ion is given by
\begin{equation}
\lb{ion_motion}
m_i \ddot{\vec{r}} = \vec F_{ij}+ \vec F_{id} + \vec F_{E}(z)+ \vec F(r, z) + \vec F_{in}
\end{equation}
where 
\begin{equation}
\lb{force_ion_ion}
\vec F_{ij}  = \sum_{i \neq j} \frac{q_{i} q_{j}}{4 \pi \epsilon_0 r_{ij}^3} \left(1+ \frac{r_{ij}}{\lambda_{De}}\right) \exp \left(\frac{- r_{ij}}{\lambda_{De}}\right) {\vec r_{ij}}
\end{equation}
is the electrostatic force between ions $ i $ and $ j $ separated by a distance $ r_{ij} $, ${\lambda_{De}} = \sqrt{\frac{\epsilon_0 k_B T_e}{n_e q_{e}^2}}$ is the electron Debye length; $ n_e $ and $ q_e $ ($ q_i $) are the number density and the charge of the electrons (ions) and $ T_e $ is the electron temperature, respectively. The force due to the dust particles arises from the Coulomb interaction between the ions and the dust, since electrons are depleted in the vicinity of the dust grain so that shielding is provided by the ions \citep{15b}
\begin{equation}
\lb{force_ion_dust}
\vec F_{id}  = \sum_{d} \frac{q_i Q_{d}}{4 \pi \epsilon_0 r_{id}^3} \vec r_{id}
\end{equation}
where the sum is over the $d$ dust particles and $r_{id} $ is the distance between an ion and a dust grain.

The force from the electric field in the sheath
\begin{equation}
\lb{force_Ez}
\vec F_{E} (z)  = q_i \vec E(z)
\end{equation}
acts in the vertical direction. In the sheath region above the lower electrode, the ions are accelerated by the vertical electric field, which is usually assumed to be linear. However, experiments conducted within a glass box placed on the lower electrode have suggested that in this case the electric field is not linear \cite{12aaa}. As such, a vertical electric field of the form
\begin{equation}
\lb{sheath_Ez}
  E(z) =  E_{0} + \alpha z + \beta z^2 
\end{equation}
is assumed, where $ E_{0} $ is a constant and $ z $ is the distance above the lower electrode. The constants $\alpha$ and $\beta$ are unknown coefficients describing the linear and quadratic contributions to the electric field, which are determined by an iterative procedure, described below.

The boundary conditions are established assuming an infinite region of homogeneously distributed ions of density $n_{i0}$ existing outside the boundaries of the simulation region. The force from these  external ions is given by
\begin{equation}
\lb{force_outside_ions}
 \vec F (r, z) = q_i \vec E_B(r, z)
\end{equation}
where $\vec E_B(r, z)$ is the electric field within a cylindrical cavity created by homogeneous Yukawa matter. The potential of uniformly distributed ions within a cylinder of height $H$ and radius $R$ interacting through a Yukawa potential is calculated numerically, then subtracted from a constant potential to obtain the potential within the cylindrical cavity. The resulting electric field $\vec E_B(r, z)$ is then calculated from the negative gradient of this potential.

The effect of random ion-neutral collisions is incorporated in the force $\vec F_{in}$. In a partially ionized plasma where the ion mean free path is smaller than the plasma screening length, ion-neutral collisions play an important role in the charging process, reducing the charge on a grain surface by a factor of 2-3  \citep{4,colla}. Collisions also act as a drag force to balance the acceleration provided by the sheath electric field, resulting in a constant drift velocity in a constant electric field. Ion-neutral collisions are included in this model using the null-collision method \citep{15d}, with collision data taken from the Phelps database (hosted by the LXcat project) \citep{15dd} for both isotropic and backscattering cross section.

The charge on a dust particle is calculated by determining the electron and ion currents to the grain. The ion current is calculated directly from the number of ions which cross the ion collection radius in a given time step, $N_d$ where the collection radius is given by
\begin{equation}
\lb{collection_radius}
b_c = a_d \left(1 - \frac{2 q_i \phi_d}{m_i V_s^2}\right)^{1/2}.
\end{equation}
In eq. (\ref{collection_radius}), $ a_{d} $ and $ \phi_d$ $ = {Q_d}/({4 \pi \epsilon_0 a_d})$ are the radius and potential of the dust grain, and $V_s$ is the characteristic velocity of the ions
\begin{equation}
\lb{char_vel}
V_s = \left(\frac{8 k_B T_i}{\pi m_i} + v_d ^2\right)^{1/2}
\end{equation}
where $ k_B $ is the Boltzmann constant, $m_i$ the ion mass, $T_i$ the ion temperature, and $v_d$  $( = M C_s)$ the drift velocity of an ion with $M$ defined as the Mach number. The ion drift velocity is measured in units of the sound speed $C_s = \sqrt{\frac{k_B T_e}{m_i}}$. 

The electron current is calculated employing orbital motion limited (OML) theory \citep{15dd1,15dd2} assuming the electrons obey a Boltzmann distribution,
\begin{equation}
\lb{electron_current}
I_e = 4 \pi q_e r_{d}^2 \sqrt{\frac{k_B T_{e}}{2 \pi m_e}} n_{e} \exp \left (\frac{- q_e \phi_d}{k_B T_{e}}\right ).
\end{equation}
The total change in charge on a dust particle in each time step, $ \Delta Q_d $, is then calculated by 
\begin{equation}
\lb{dust_charge}
\Delta Q_d = I_e \Delta t_i + q_i N_d.
\end{equation}

\begin{figure}[t]
\includegraphics[width=1.01\columnwidth]{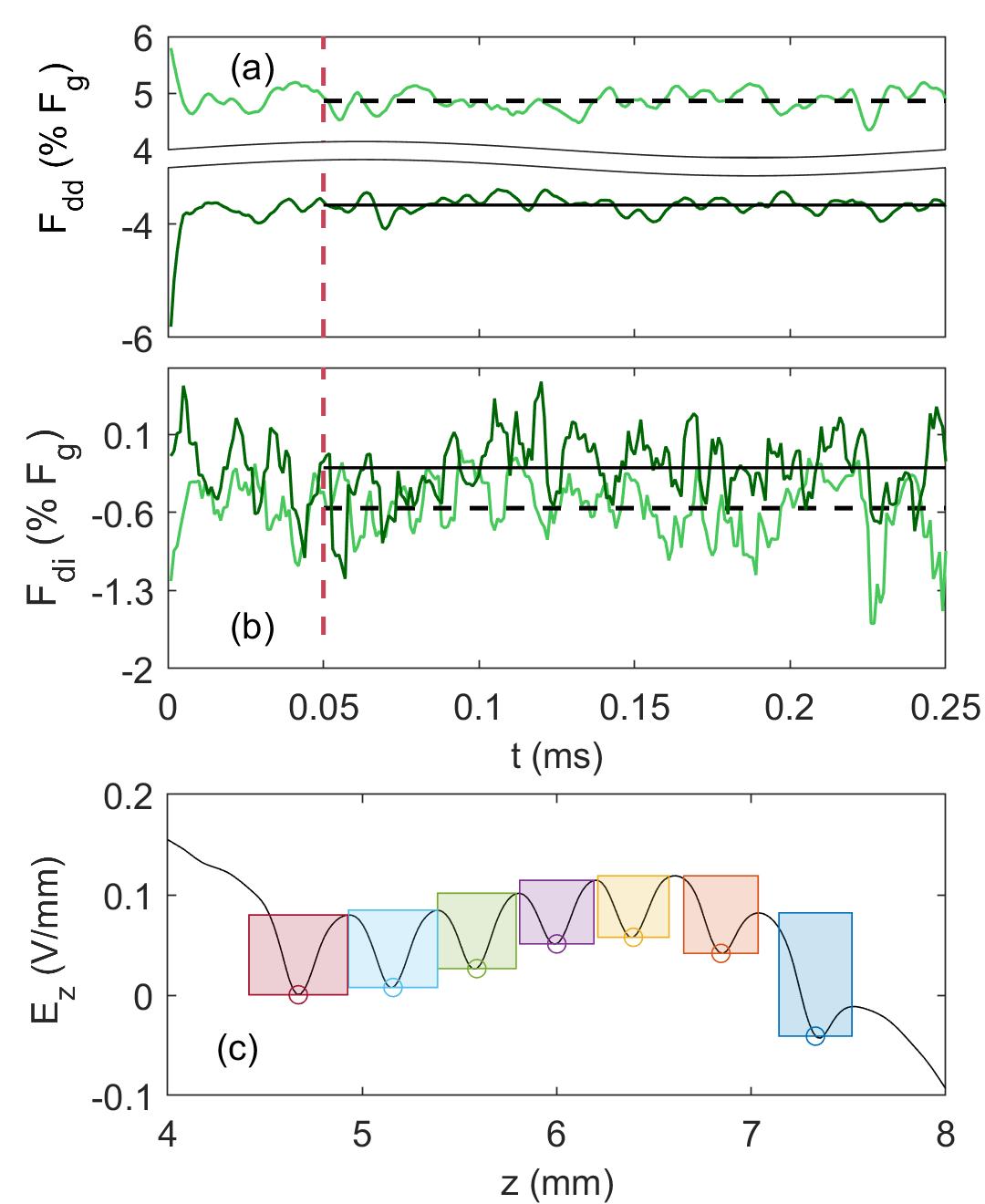}
\centering
\caption{(Color online) (a,b) Forces acting on dust from the other dust grains and the ions, respectively, for the top (light green) and bottom (dark green) grains as a function of time. For clarity, the data shown is a moving average over $10$ dust time steps. After the system reaches equilibrium (indicated by the dashed red line), quantities are averaged over the remaining time (indicated by the horizontal lines). (c) Vertical electric field derived from the ion potential.  Each shaded box shows the extent of the spherical regions used to calculate the dust-ion force as shown in (b). }
\label{fig2}
\end{figure}

Although dust grains are held at fixed positions in this simulation, the forces acting on the dust grains (defined in Eq. \ref{dust_force}) are calculated at each time step. The (Coulomb) interaction between any two grains with vertical separation $ z_{jk} = z_j - z_k $ is given by
\begin{equation}
\lb{force_dust_dust}
\vec F_{dd} = \sum_{j \neq k} \frac{Q_{d_j} Q_{d_k} }{4 \pi \epsilon_0 z_{jk}^3} {\vec z_{jk}};
\end{equation}
where shielding is not included since we explicitly model the force from the ions which includes shielding effects \citep{15b}. The forces calculated for the top and bottom grains are illustrated in Fig. \ref{fig2}a.  

The force from the ions not only provides the shielding for the interaction between the grains, but also encompasses the ion (orbit) drag exerted by the flowing ions as well as the ion wake forces acting on the downstream grains. This force is given by

\begin{equation}
\lb{force_dust_ion}
\vec F_{di}  = \sum_i \frac{q_i Q_{d}}{4 \pi \epsilon_0 r_{di}^3} \left(1+ \frac{r_{di}}{\lambda_{De}}\right) \exp \left(\frac{- r_{di}}{\lambda_{De}}\right) {\vec r_{di} + Q_d \vec{E}_{out}} 
\end{equation}
where the sum is over the $i$ ions inside the simulation and $\vec{E}_{out}$ is the electric field from the homogeneously distributed ions outside the simulation region. The time-evolved trace of this force is very noisy as ions which pass close to a dust grain exert a large force over a very short period of time. 

In this study, this force is smoothed by integrating the Maxwell stress tensor over the surface of a sphere centered on each dust grain \cite{Hutchinsonppcf05},
\begin{equation}
\label{stress_tensor}
\vec F_{di} = \int_S \epsilon_0 \left(\frac{1}{2}E^2 \openone - \vec{E}\vec{E}\right) d\vec{S}
\end{equation}
which gives the net electric force on all particles inside the surface.  The time evolved trace for this force is illustrated for the top and bottom grains in Fig. \ref{fig2}b. The stress tensor is derived from finite differences of the potential calculated by summing the contributions from all of the Yukawa ions in the simulation and the assumed homogeneous distribution of Yukawa ions outside the simulation region, averaged over ten dust time steps.

The radius of the spherical surface for each dust grain is determined by examining the vertical component of the electric field derived from the potential along the line of the dust grains.  As shown in Fig. \ref{fig2}c, each dust grain is situated within a “well” in the electric field. The radius of the well for each dust grain is determined by the minimum distance from the dust grain to a local maximum in the electric field, as indicated by the shaded boxes in Fig. \ref{fig2}c.  The radius of each well is approximately equal to the cutoff radius inside of which the dominant radial dependence of $\phi$ is the $1/r$ Coulombic potential of the dust grain \cite{Hutchinsonpop13}.  Taking the cutoff radius to be the radius of a cloud of ions which contains an ion charge equal to $-Q_{d}/2$, $r_c \approx 1.2 (r_d T_i/\lambda_{De} T_e)^{1/5}\lambda_{De}$ \cite{Hutchinsonpop13}.  At this distance, the contribution from the ion momentum flux (which is almost, but not quite the same as the ion collection drag) is much smaller than the contribution from the Maxwell stress, and are neglected.

The simulation is advanced through a leapfrog integration scheme until it comes to a steady state. For the case at hand, the simulation runs for 250,000 time steps ($\approx 83$ $\tau_{i}$, where $\tau_{i} = 2\pi \sqrt{\epsilon_0 m_i/n_i q_e^{2}}$ is the ion plasma period). The first 50,000 steps ensure the system is in equilibrium, with the quantities of interest averaged over the remaining 200,000 steps.
~~~~~~~~~~~~~~~~~~~~~\
~~~~~~~~~~~~~~~
~~~~~~~~~~~~~~~~~~
\section*{IV. Method and Results}

As the plasma density in the sheath is decreased from that in the bulk, the initial number density of the plasma in the simulation is taken to be $ n_{e0} = n_{i0} = 1\times 10^{14}$ $m^{-3}$ \citep{15b,15bbb} and $T_e \approx 58022$ $K$ ($5$ $eV$). Ions are taken to be at room temperature as they thermalize with the neutral gas $T_i=T_g \approx 290$ $K$. The calculated value of $\lambda_{De}$ is $1.65$ $mm$ using the above parameters. The simulation is carried out in a cylinder of radius $R = 1.25$ $\lambda_{De} $, and total height $H = 5.0$ $\lambda_{De}$. The height of the cylinder is chosen such that the dust grains are at least one to two Debye lengths from the top and bottom boundaries. 

A fluid model of the plasma within the CASPER GEC cell gives an estimate for the sheath electric field $E(z)$ ranging from $-3.5$ to $-1.25$ $V/mm$ within the sheath \citep{15dc1}. Previous experiments with these conditions yield an estimated grain charge of $\approx 13500$  $e^{-}$ \citep{12aa}. Accordingly, a range of initial values for $Q_d$ and $E(z)$ were chosen in order to allow for different ion flow regimes such that $Q_d E(z) = m_{d} g$. According to sheath theory, ions enter the sheath at the Bohm velocity, $C_s$ and are accelerated towards the lower electrode \citep{15ea}. However, measurement of the horizontal restoring force acting on a grain in the wake of another grain \cite{12aaa} combined with results from a previous numerical model of the ion wake \cite{15b} suggests an ion flow speed $M<0.4$ inside the lower portion of the glass box. Therefore, as shown in Table \ref{table:1}, three values of the ion flow speed were chosen for the initial iteration $M= 0.46$, $0.92$, $1.4$ in order to investigate subsonic-supersonic ion flow. The corresponding (constant) electric fields $E_0^{(1)}$ were estimated from \citep{15e} 

\begin{equation}
\lb{drift_vel}
v_d = \left[\left[\left(\frac{3  k_B T_i}{m_i}\right)^2 + \frac{3}{K^2 m_i} \left(\frac{E}{N}\right)^2 \right]^{\frac{1}{2}} - \frac{3  k_B T_i}{m_i}\right]^{\frac{1}{2}}
\end{equation}
where $K= [{8}/{3q_e}]\left({m_i}/{\pi}\right)^{1/2}\pi b_{ex}^2$, and $N$ is the number density of the neutral gas. The impact parameter for charge exchange, $b_{ex}$, is a function of the energy of the ions, $\epsilon$, and the polarization potential, determined from $b_{ex} \left({\epsilon}/{4 a}\right)^{1/4} \equiv A \epsilon^{1/4}$, where the constant $A$ is determined from experiment and $a = ({\alpha_d q_e ^2})/({2 (4 \pi \epsilon_0)^2})$ where $\alpha_d$ is the polarizability. For argon, $\alpha_d = 1.642 \times 10^{-30}$ $m^{-3}$ and $A = 2.6$ $eV^{-1/4}$ \citep{15e}.

\begingroup
\setlength{\tabcolsep}{0.25em} 
\renewcommand{\arraystretch}{1.2} 
\begin{table}[ht]
\def\arraystretch{1.25}
\caption{Simulation parameters for three cases, giving the initial values for a constant electric field (Eq. \ref{drift_vel}) as well as the derived values for the linear and quadratic fields after the fifth iteration. The drift velocity is given at the position of the top grain $z=7.33$ $mm$. } 
\centering 
\begin{tabular}{l l l l} 
\hline\hline\
$E_0^{(1)}$ & Case I & Case II & Case III\\
$E_0$ ($V/mm)$ & -4.790  & -16.00 & -32.50 \\
$v_d$ ($M$) & 0.460 & 0.920 & 1.400 \\
\hline 
$E_{lin}^{(5)} = E_0 + \alpha z$  \\
$E_0$ ($V/mm)$ & -4.297 & -4.092  &  -4.299 \\
$\alpha$ ($V/mm^2)$& 0.239 & 0.206  & 0.240  \\
$v_d$ ($M$) & 0.319 & 0.322 & 0.320\\
\hline 
$E_{quad}^{(5)} = E_0 + \alpha z + \beta z^2$  \\
$E_0$ ($V/mm)$& 0.646 & -0.798  &  -0.720\\
$\alpha$ ($V/mm^2)$& -1.500 & -1.001 & -1.000 \\
$\beta$ ($V/mm^3)$& 0.150 & 0.108 & 0.106\\
$v_d$ ($M$) & 0.302  & 0.300 & 0.307\\
\hline \hline 
\end{tabular}
\label{table:1} 
\end{table}
\endgroup

\begin{figure}
\includegraphics[width=\columnwidth]{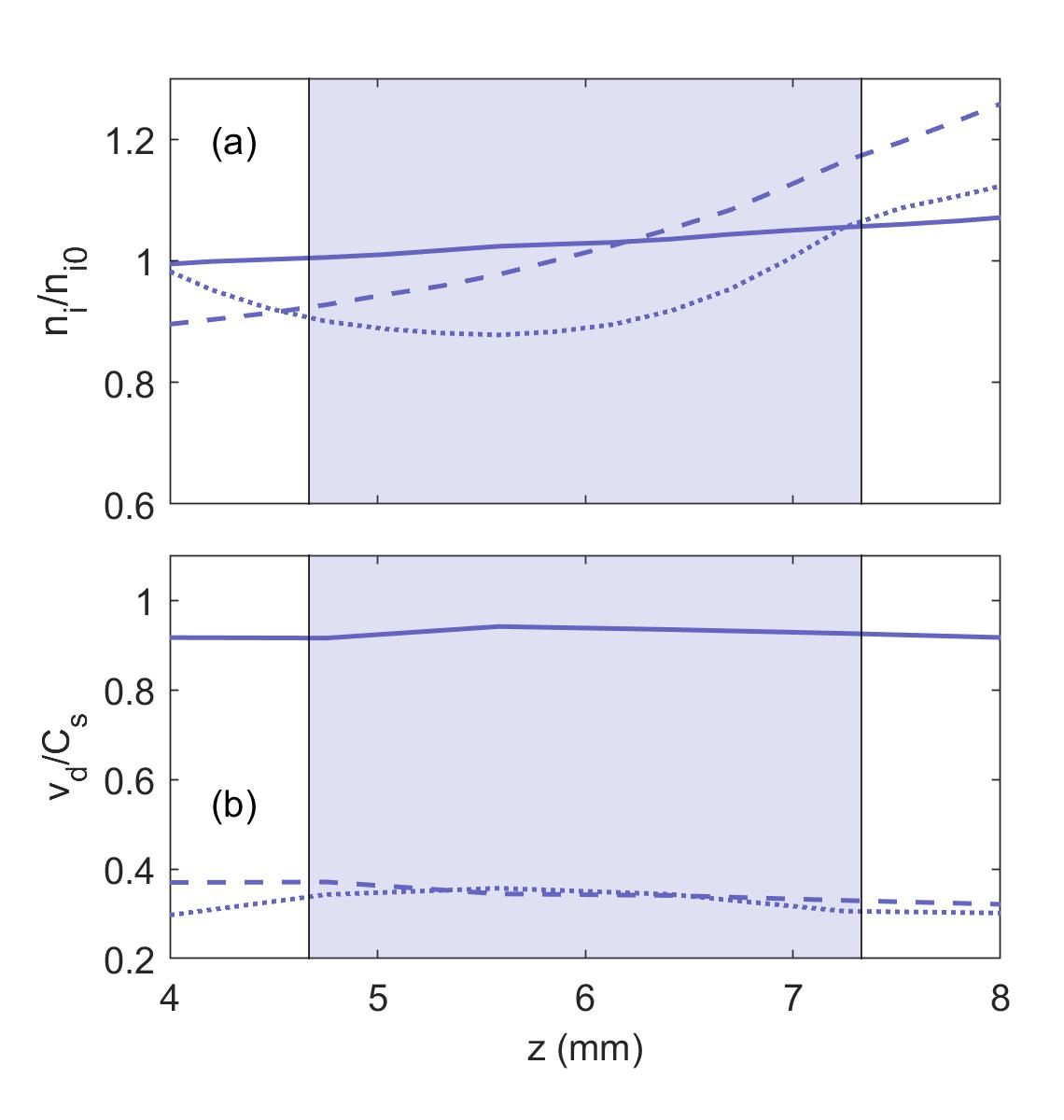}
\centering
\caption{(Color online) (a) Ion density (normalized by background ion density), and (b) drift velocity (normalized by sound speed) within the simulation region for constant (solid line), linear (dashed line), and quadratic (dotted line) electric fields. The shaded region indicates the vertical extent of the seven dust grains in the chain, measured relative to the lower electrode. The electric field and flow velocity are provided in Table \ref{table:1}, Case II. } 
\label{fig3}
\end{figure}

\begin{figure}
\includegraphics[width=\columnwidth]{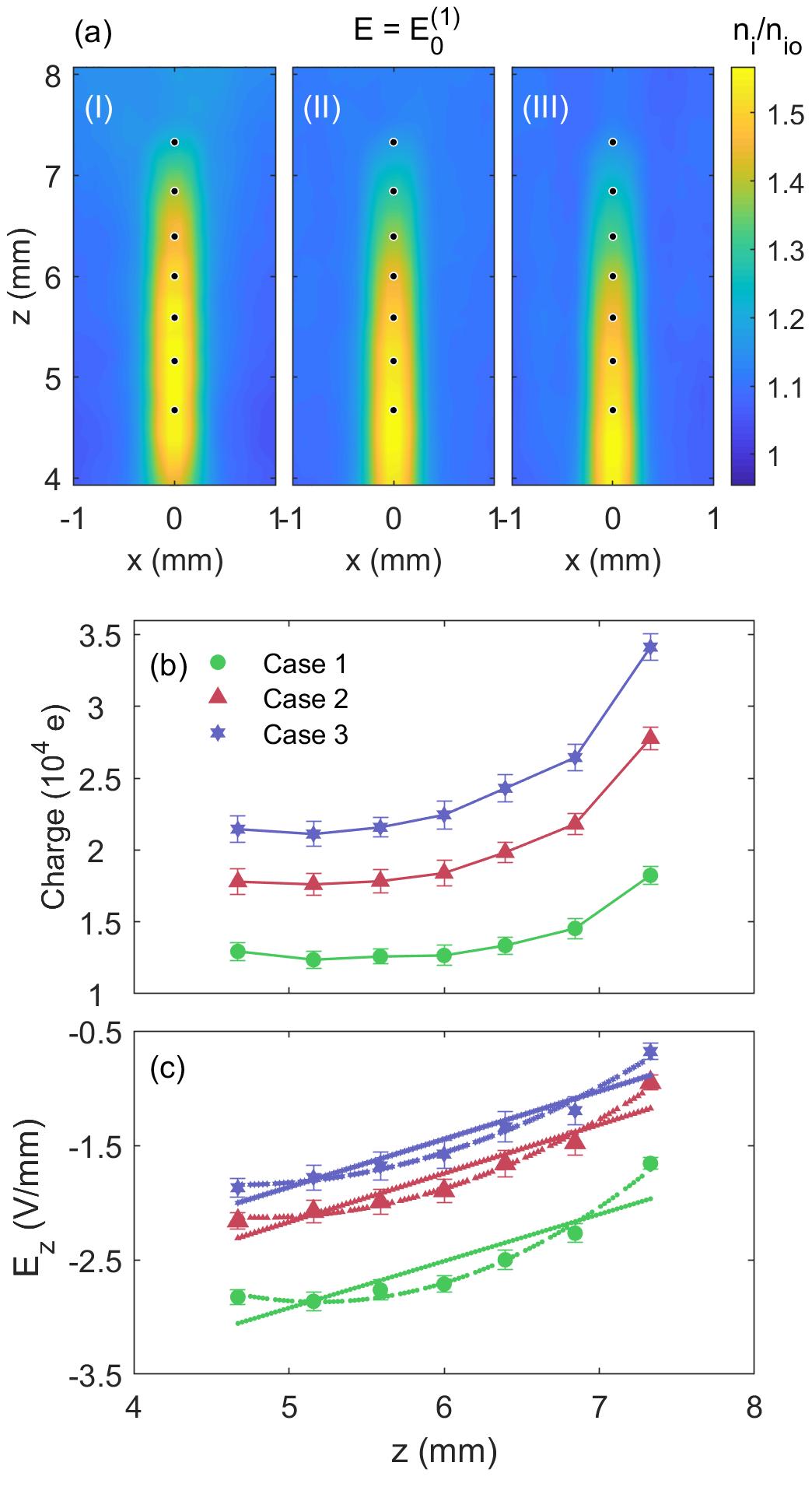}
\centering
\caption{(Color online) Results from initial simulations assuming a constant electric field $E_0^{(1)}$. Cases I, II, and III correspond to the three cases listed in Table \ref{table:1}. (a) Ion density distribution. (b) Equilibrium charge on the dust. The error bars indicate the standard deviation of the calculated values, with the lines serving to guide the eye. (c) Calculated electric field at the position of each dust grain.  Linear (solid lines) and quadratic (dashed lines) fits are used to determine the coefficients $\alpha$ and $\beta$. Note that the initial values of $E_0$ are outside the range of the plot.  }
\label{fig4}
\end{figure}
Once the ions and dust charges reach a steady state, the average charge $Q_d$ and forces $\vec F_{dd}$ and $ \vec F_{di}$ are determined from the simulation data as illustrated in Fig. \ref{fig2} using the data for the top and bottom grains. For clarity, the data shown at each time step are expressed as a moving average over 10 time steps to reduce the large stochastic fluctuations and better indicate the difference in forces acting on the two grains. The dust-dust force is directed downward on the bottom grain, as it is repelled by the upper grains, but its magnitude is reduced as it is located in a region of high ion density and has a smaller charge. The downward directed dust-ion force is an indicative of the wake field formed downstream of the grains. The resultant difference in ion density provides a greater dust-ion force on the top grain.

In the first iteration, it is assumed that the electric field is constant ($E_0^{(1)}$, where the superscript refers to the first iteration). In the absence of dust, the ion density and flow speed are relatively constant throughout the cylinder, as shown in Fig. \ref{fig3} for $v_d = 0.92 M$ (Case II). After the first iteration, the time-averaged values of the forces as illustrated in Fig. \ref{fig2} are used to solve for the electric field at the position of each grain by applying the force balance equation (Eq. \ref{dust_force}).  

 The ion density distribution near each grain in the chain, the charge on each grain, and the calculated vertical electric field are shown in Fig. \ref{fig4} for $E_{0}^{(1)}$ for each of the three cases. As shown in Fig. \ref{fig4}a, as the ion flow speed increases, the position of the maximum ion focus shifts downward. Differences in local ion density also affect the equilibrium charge on the grains, with the lower grains being decharged relative to the upper grain (Fig. \ref{fig4}b). The calculated electric field in each of the cases (Fig. \ref{fig4}c) was fit with both a linear and quadratic function of $z$ in the region spanned by the dust chain, to determine the coefficients $\alpha$ and $\beta$, and these fields, $E_{lin}^{(2)}$ and $E_{quad}^{(2)}$, respectively, were used in the next iteration.

\begin{figure}
\centering
\includegraphics[width=1.01\columnwidth]{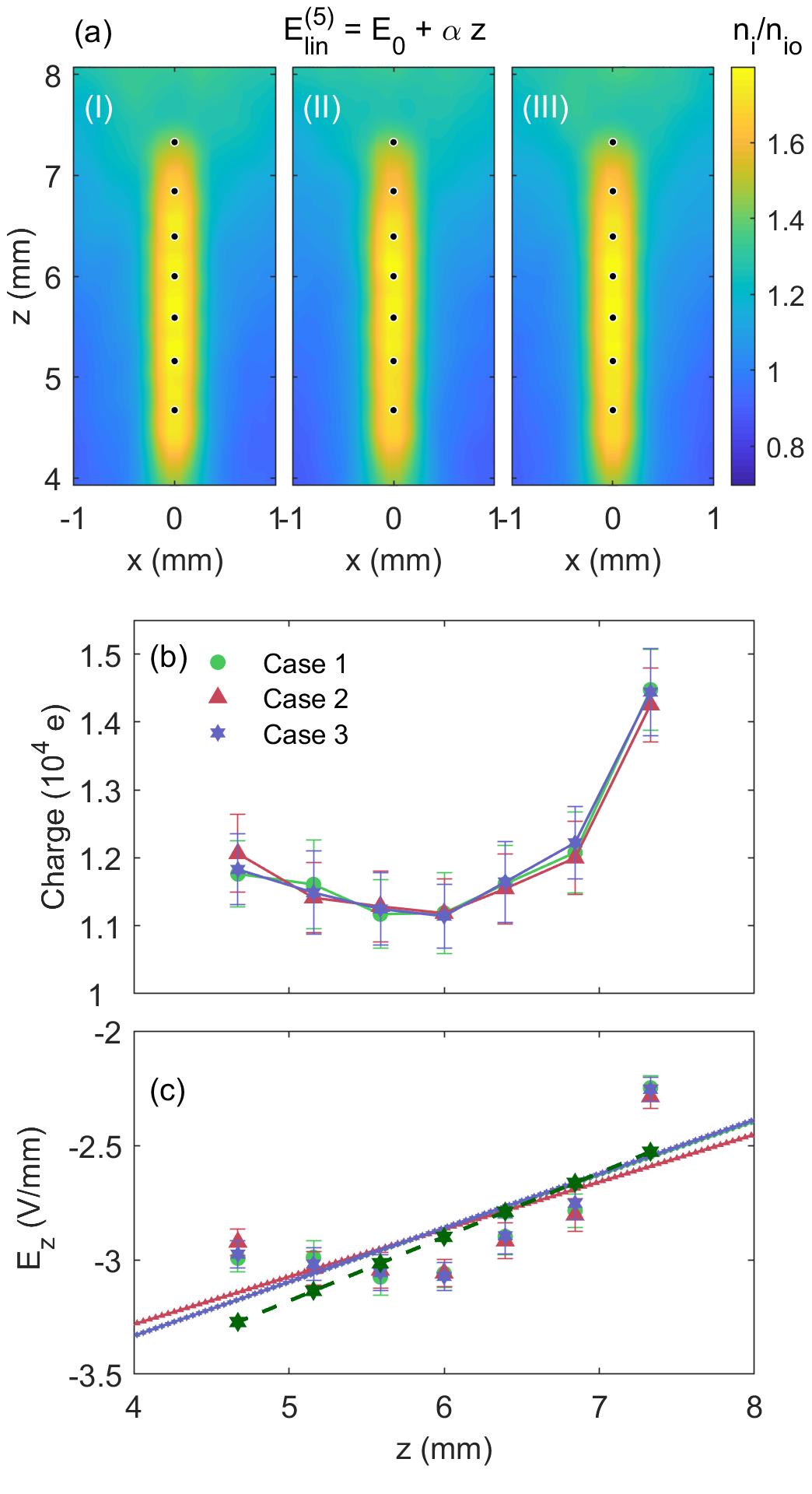}
\centering
\caption{(Color online) Results from fifth iterations assuming a linear applied electric field. Cases I, II, and III correspond to the three cases listed in Table \ref{table:1}. (a) Ion density distribution. (b) Equilibrium charge on the dust. The error bars indicate the standard deviation of the calculated values, with the lines serving to guide the eye. (c) Calculated electric field at the position of each dust grain. The green dashed line shows the value of $E^{(4)}(z)$ used as input (only one line is shown here for illustration).}
\label{fig5}
\end{figure}
This new electric field was then used to estimate the ion flow speed at the upper dust grain's position ($z_u$). $E(z)$ is taken as a piecewise function, and in the region above the dust is set to $E(z) = E (z_u)$ to ensure that $E(z)$ does not exceed zero (its value in the bulk plasma) and that ions inserted at the top of the box are accelerated downward and reach the appropriate drift velocity by the time they reach the dust grains. Note that with a varying electric field, the number density (Fig. \ref{fig3}a) and drift velocity (Fig. \ref{fig3}b) measured without dust present vary by $13-21$ $\%$ throughout the region occupied by the dust. 

After the simulation reaches equilibrium, the values of $Q_d$ and all forces are calculated as before and the electric field $\vec E^{(2)}(z)$ is subsequently obtained. The method is then iterated until the cases converge and the initial and final electric fields agree, as shown in Figs. \ref{fig5} and 6 for assumed linear and quadratic fields, respectively, and the total mean force acting on the dust chain is minimized. The total mean force is calculated from the mean of the absolute value of total force acting on each dust grain, shown in Fig. \ref{fig7}. 

The final values of $\alpha$, $\beta$ and $v_d$ are given in Table \ref{table:1}. The final average dust charge, along with the total forces acting on each grain for $E_{lin}^{(5)}$ and $E_{quad}^{(5)}$ are given in Table \ref{table:2} for Case I, which provides the best fit, though the results for the other two Cases are similar. As shown, the quadratic electric field gives the minimum force balance for the grains. The error in  $E_{quad}^{(5)}$, expressed as a percent difference from that needed to exactly balance the forces on each grain is listed in Table \ref{table:3}. In general, the percent error in the calculated electric field is 2-3$\%$, with the maximum error of 4.55$\%$ found for the bottom grain.

\begin{figure}
\includegraphics[width=\columnwidth]{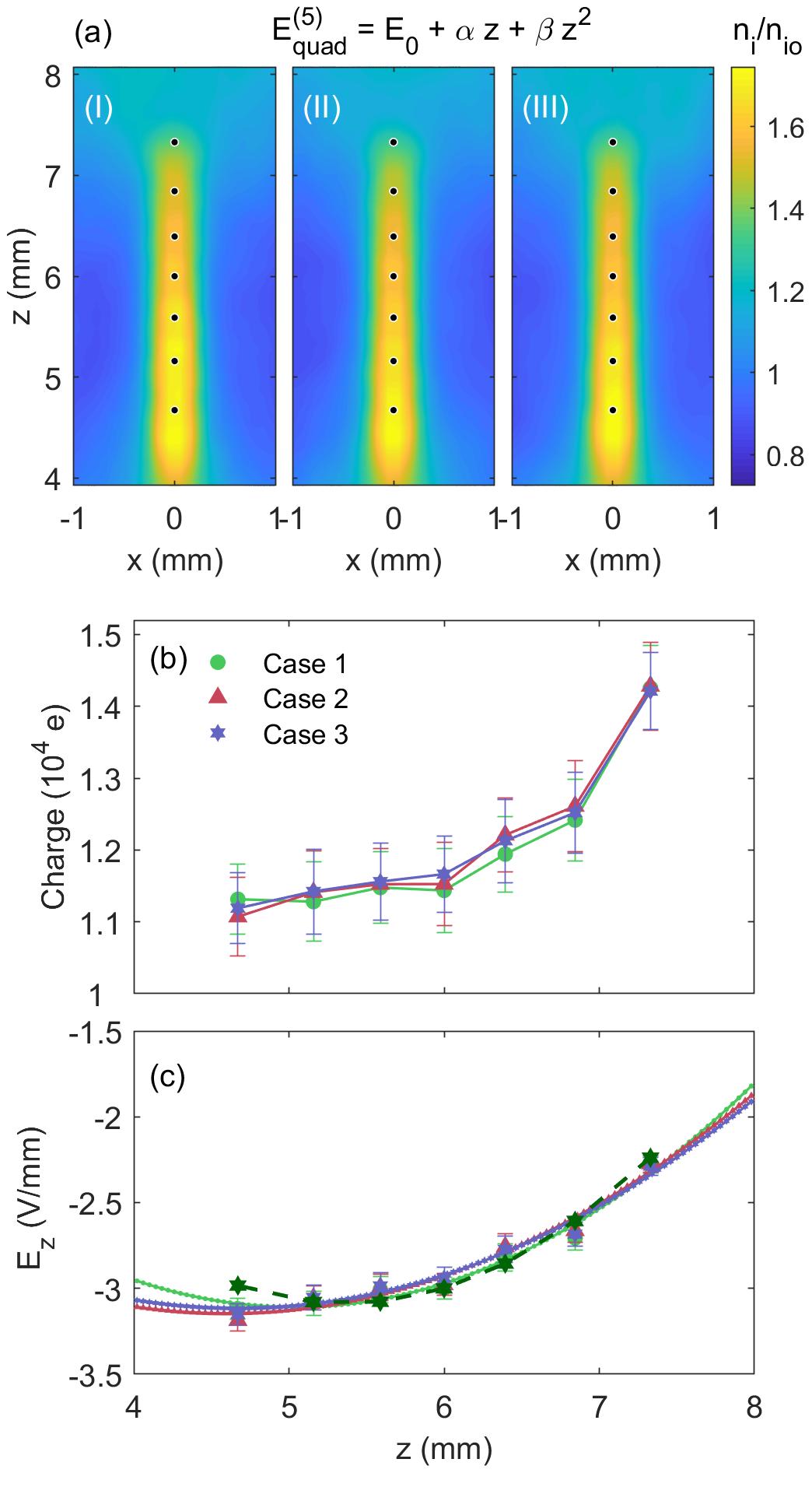}
\centering
\caption{(Color online) Results from fifth iterations assuming a quadratic applied electric field. Cases I, II, and III correspond to the three cases listed in Table \ref{table:1}. (a) Ion density distribution. (b) Equilibrium charge on the dust. The error bars indicate the standard deviation of the calculated values, with the lines serving to guide the eye. (c) Calculated electric field at the position of each dust grain. The green dashed line shows the value of $E^{(4)}(z)$ used as input (only one line is shown here for illustration). }
\label{fig6}
\end{figure}
\begin{figure}[t]
\includegraphics[width=1.02\columnwidth]{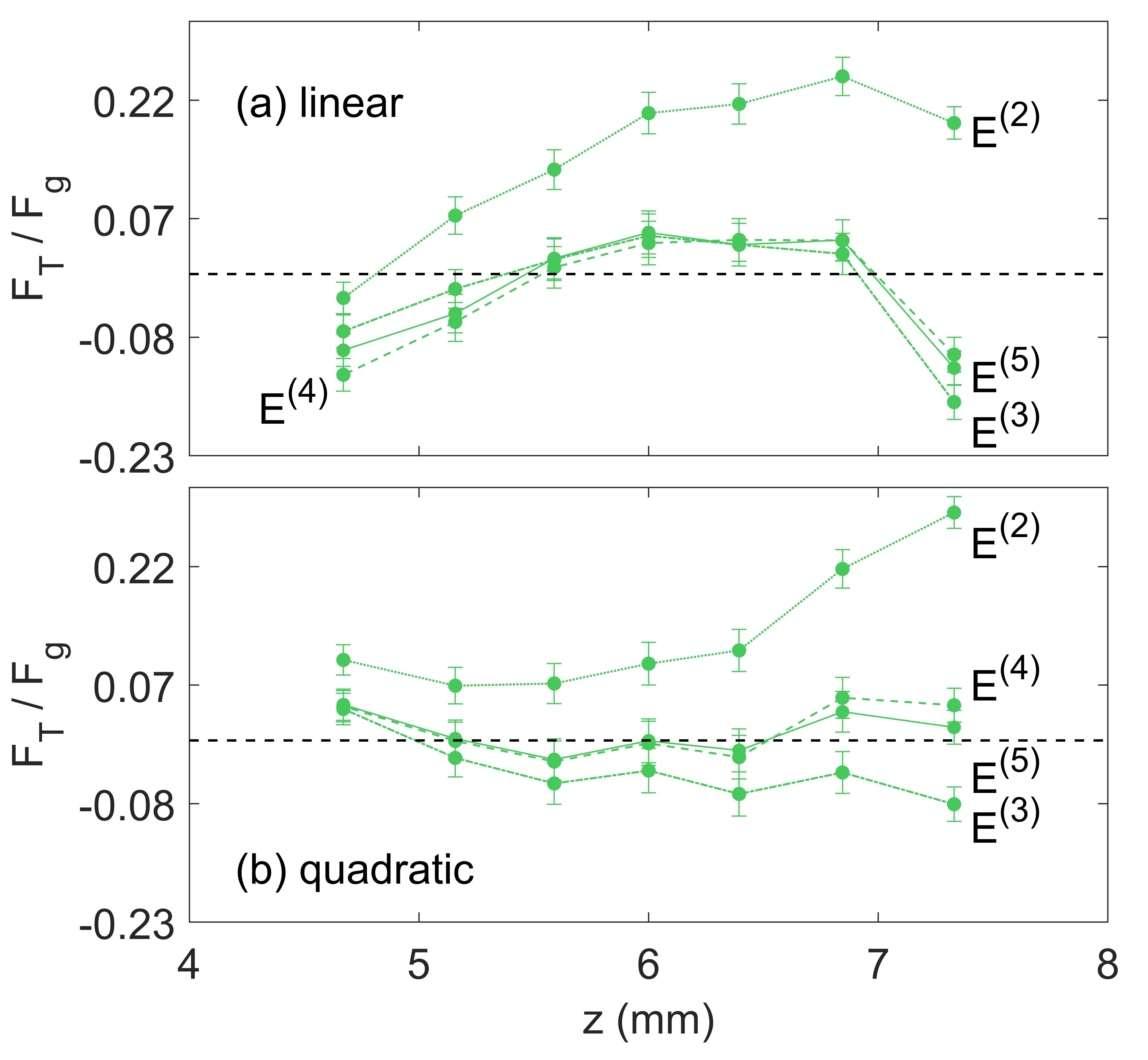}
\centering
\caption{(Color online) Total force acting on each grain (normalized by gravity) for Case I with the assumption of (a) linear and (b) quadratic sheath electric field. Dotted (second iteration), dashed-dotted (third iteration), dashed (fourth iteration), and solid (fifth iteration) lines indicate different iterations. }
\label{fig7}
\end{figure}

\section*{V. DISCUSSION}

All three Cases converge to almost the same values of dust charge and electric field, under the assumption of either a linear or quadratic electric field. The final results for both linear and quadratic fits yield similar magnitudes for the electric field with a subsonic ion flow velocity $v_d \approx 0.3$. The calculated charge for the top grain, $Q_d = 14,260-14,480$ $e^{-}$, is within 7$\%$ of the charge measured experimentally under similar plasma conditions \citep{12aa}. After the fifth iteration, the mean force acting on the grains was $F_{mean}/m_dg$ =  1.96$\%$ for the quadratic electric field, while that calculated for the linear electric field was 5.97$\%$ (Table \ref{table:2}). These results favoring the quadratic electric field corroborate the results from a different numerical model, which calculated the electric field at the center of the box by adding the electric field from the charge collected on the walls of the box to the sheath electric field determined from a plasma fluid model \citep{15dc1,15bb}.

The subsonic ion flow coupled with the ion wake field change the ion density downstream of the grains, causing variations in the equilibrium dust charge along the length of the chain. The top grain always has the largest charge, and the results from the first iteration with a constant ion flow speed show that the charge ratio of the top two grains decreases as the ion flow speed increases (Fig. \ref{fig4}b). The resulting charge ratios $Q_{d_{2}}/Q_{d_{1}}= 0.797, 0.785$, and $0.775$ for $M = 0.46$, $M = 0.92$, and $M = 1.4$, respectively, are in agreement with results from previous numerical simulations for subsonic, sonic, and supersonic ion flow using a PIC code which also found that the charge ratio decreases as the ion flow increases \cite{15,15ec}. As indicated in Table \ref{table:2}, the best-fit data from the final iteration of this model give values for $Q_{d_{2}}/Q_{d_{1}}= 0.834$ for a linear electric field and $0.871$ for a quadratic electric field. Previous experimental measurements of this charge ratio for a vertically aligned particle pair yielded values of $0.78$ - $0.80$ \cite{15g, 15h}. The larger charge ratio found in this simulation may be due to the addition of the downstream dust grains changing the ion focus, or that the plasma conditions in this experiment result in a smaller ion flow. It is also important to note that the quadratic electric field results in a dust particle charge which decreases almost monotonically down the chain, while a linear electric field model results in the minimum grain charge for the particles located in the center of the chain.

\begin{figure}[t]
\includegraphics[width=0.9\columnwidth]{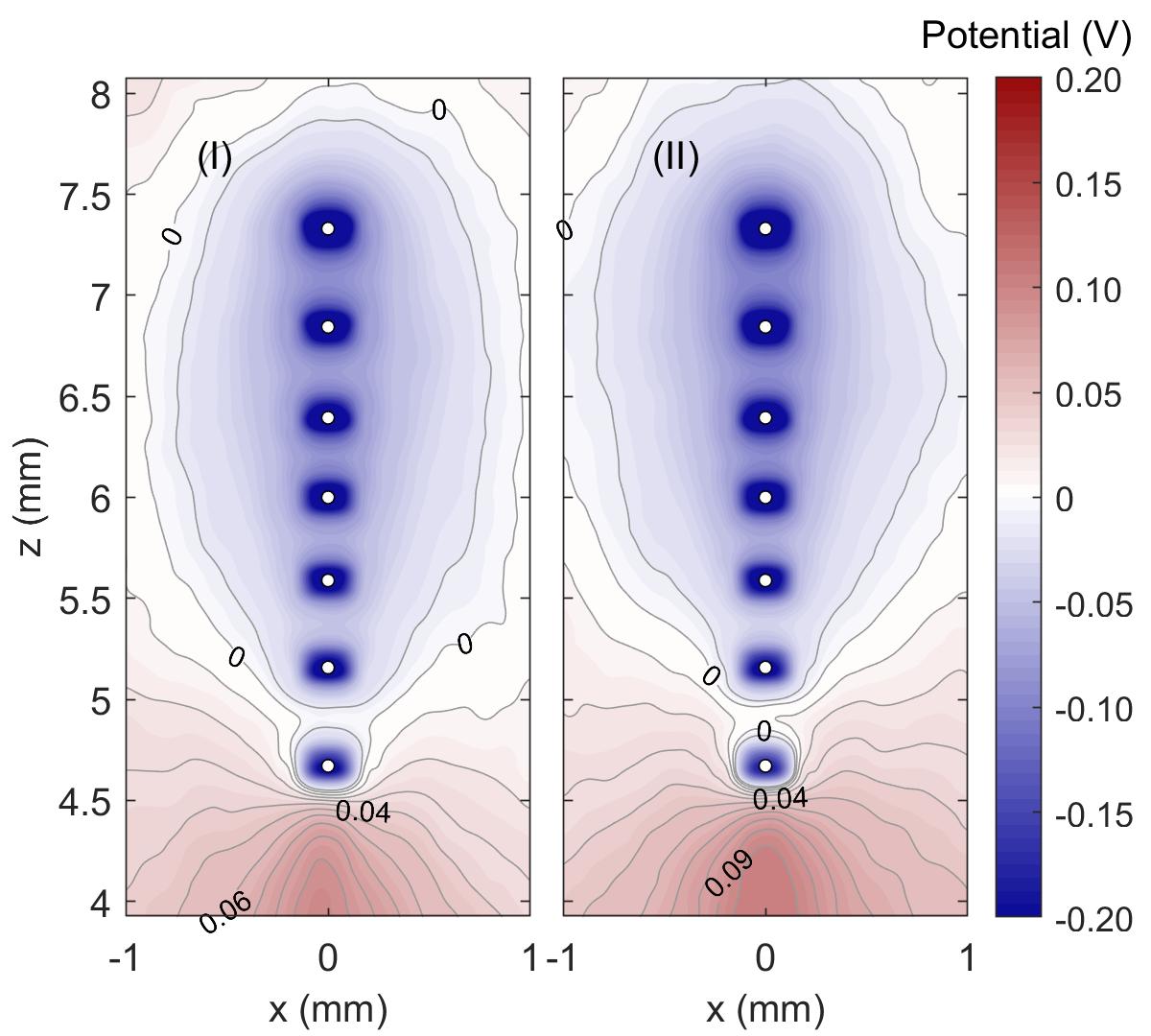}
\centering
\caption{(Color online) Combined potential of ions and dust grains for (I) $E_{lin}^{(5)}$ and (II) $E_{quad}^{(5)}$ (Case I). The contour lines indicate levels from $-0.01:0.01:0.20$ $V$. }
\label{fig8}
\end{figure}

Another difference which arises depending on the assumption of a linear or quadratic sheath electric field is the location of the maximum ion density within the wake field (Figs. 5a and 6a). The maximum ion density is found to be in the middle of the chain for $E_{lin}^{(5)}$, while for $E_{quad}^{(5)}$ it is greatest at the bottom of the chain. Figure \ref{fig8} shows the electric potential calculated for both the ions and dust using $E_{lin}^{(5)}$ and $E_{quad}^{(5)}$ of Case I, measured relative to the background ion potential calculated for these conditions without the dust present. Notice that the enhanced ion density near the bottom of the chain produces a maximum potential below the bottom grain, which is stronger in the case of the quadratic electric field. Previous PIC simulations of subsonic ion flow \cite{15i,15j} have shown that there are regions with slightly positive potential between the negative dust grains, which contribute to the stability of the vertical alignment of grains. Such positive regions are seen here between grains 6 and 7 in the case of the quadratic electric field. 

Figure 9 compares simulation results to experimentally obtained values for ${Q_d E}/{m_d g}$ at vertical positions within the box \citep{12aaa}. In this case, the experimental values were determined by a free-fall technique where the charge and dynamics of isolated dust grains are only influenced by the local equilibrium plasma parameters. Note that while the experimental values of ${Q_d E}/{m_d g}$ for most of the particles fall within the error bars of the simulation results (and vice versa) for both the linear and quadratic electric field, the general trend for ${Q_d E}/{m_d g}$ tends to be less steep than that observed in the numerical model. The major contribution to this deviation is likely due to the fact that in the simulation the ion density decreases towards the lower electrode, whereas the electron density is held constant throughout the simulation region, thus causing overcharging of the grains near the bottom of the chain. A decrease in $n_e$ would result in a decreased dust charge at the lower end of the chain. Additional modelling by including the variation of the electron density along the vertical direction is needed to address the error obtained in the calculated electric field for the lower particles in the chain. This will be the subject of future work.

\begin{figure}[t]
\includegraphics[width=1.05\columnwidth]{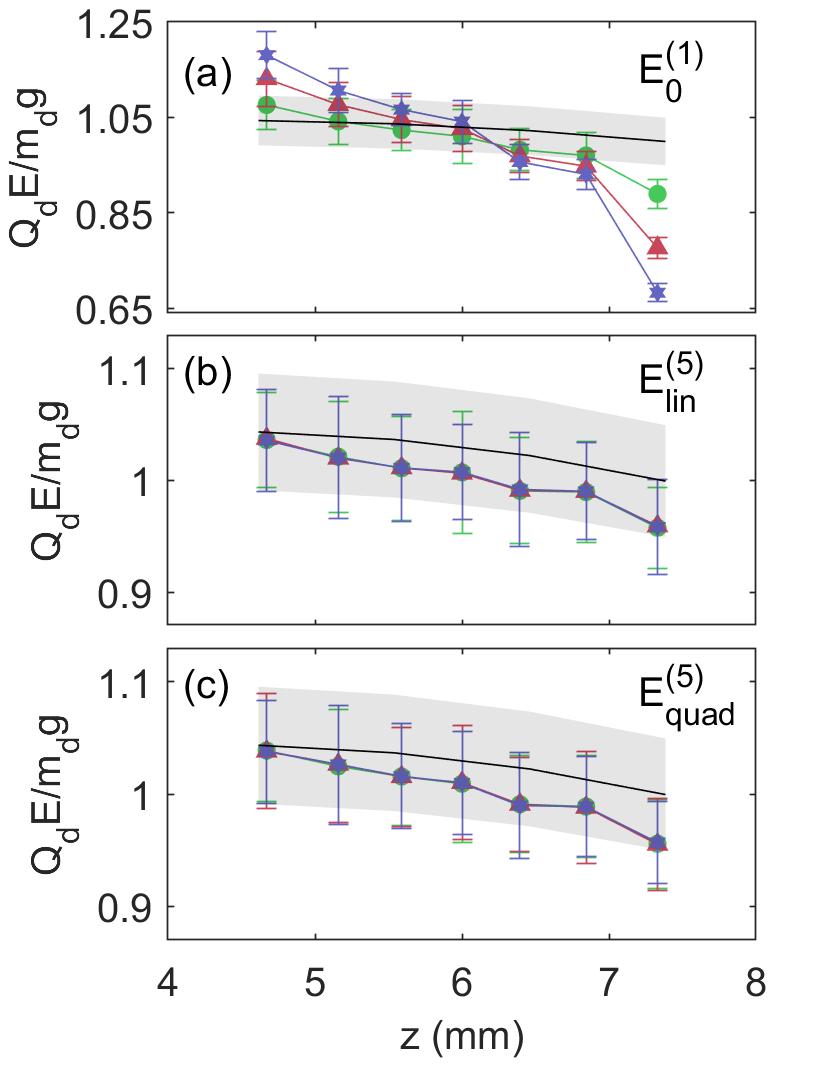}
\centering
\caption{(Color online) Electrostatic force normalized by gravity for seven dust grains as a function of vertical distance for (a) the initial constant and (b, c) final linear and quadratic electric fields. Error bars show the standard deviation from the time-averaged force. The black line indicates experimentally measured values with a 5\% error indicated by the shaded region. The green circles indicate data for Case I, red triangles and blue stars indicate data for Case II and Case III,
respectively as described in the text and Table \ref{table:1}. }
\label{fig9}
\end{figure}
\begingroup
\setlength{\tabcolsep}{0.15 pt} 
\renewcommand{\arraystretch}{1.0} 
\begin{table}[ht]
\def\arraystretch{1.2}
\caption{Values of the dust charge and total normalized force for $E_{lin}^{(5)}$ and $E_{quad}^{(5)} $.} 
\centering 
\begin{tabular}{c c c c c} 
\hline\hline 
 & \multicolumn{1}{c} {\centering $Q_{lin}^{(5)}$ ($10^{4}$ $e)$}
& \multicolumn{1}{c}{\centering ${F_{T_{lin}}^{(5)}}/{F_g} (\%)$}
& \multicolumn{1}{c}{\centering $Q_{quad}^{(5)}$($10^{4}$ $e)$}
& \multicolumn{1}{c}{\centering ${F_{T_{quad}}^{(5)}}/{F_g} (\%)$}\\
\hline
${d_1}$ & 1.448 & -11.9& 1.426& 1.65\\ 
${d_2}$ & 1.208 & 4.30& 1.242& 3.63\\
${d_3}$ &1.162 & 3.72& 1.194& -1.26\\
${d_4}$ &1.119 & 5.25& 1.144& -0.05\\
${d_5}$ &1.117 & 1.97& 1.148& -2.41\\
${d_6}$ &1.161 & -4.99& 1.128& 0.23\\
${d_7}$ &1.176 & -9.65& 1.131& 4.52\\
\hline
$<|{F_T}|>$ &  & 5.97 & & 1.96\\
\hline \hline 
\end{tabular}
\label{table:2} 
\end{table}
\endgroup

\begingroup
\setlength{\tabcolsep}{1.0 pt} 
\renewcommand{\arraystretch}{1.0} 
\begin{table}[ht]
\def\arraystretch{1.2}
\caption{Percent error in the calculated electric field $E_{quad}^{(5)} $ (Case I).} 
\centering 
\begin{tabular}{c c c c c c c c c} 
\hline\hline 
 & \multicolumn{1}{c} {\centering ${d_1}$}
& \multicolumn{1}{c}{\centering ${d_2}$}
& \multicolumn{1}{c}{\centering ${d_3}$}
& \multicolumn{1}{c}{\centering ${d_4}$}
& \multicolumn{1}{c}{\centering ${d_5}$}
& \multicolumn{1}{c}{\centering ${d_6}$}
& \multicolumn{1}{c}{\centering ${d_7}$}
& \multicolumn{1}{c}{\centering ${\langle |\Delta E_{z,i}|\rangle}$}\\
\hline
$\Delta E_{z,i}/E(z_i)$ & -1.76 & -3.8& 1.25& 0.05& 2.31& -0.21& -4.55& 1.99\\ 
\hline  
\end{tabular}
\label{table:3} 
\end{table}
\endgroup

\section*{VI. CONCLUSIONS}

A numerical model of ions flowing past charged dust grains was used to determine the dust charge, ion density, electric field in the region spanned by the dust chain, and ion flow velocity for a vertical chain of seven grains formed within a glass box placed on the lower electrode in a GEC rf reference cell, based on the experimentally measured equilibrium positions of the grains. The model was initialized starting with three different values of constant ion flow velocity and associated constant electric fields producing the ion flow. The resulting dust charges and ion densities were used to determine sheath electric fields which led to stable structures, and the model then iterated assuming either a linear or quadratic fit to the calculated electric field.  

The results converge within 1$\%$ for all three cases, with the linear and quadratic fits giving similar results for the dust charge, electric field, and ion flow velocity, though the total force balance indicates that a quadratic fit to the electric field provides the best results.  The ion flow velocity was found to be subsonic throughout the region spanned by the dust chain, with downstream grains decharged by $17-20$ $\%$ compared to the top grain.  As a flow velocity less than the Bohm velocity is characteristic of the presheath region, the subsonic ion flow inside the glass box obtained from this study indicates that the glass box extends the presheath region.

The varying dust charge and ion density along the length of the chain are responsible for the alignment and stability of the vertical chain structures observed in ground-based experiments \citep{12aa,12aaa,12n,3,15jj}. Stable particle chains have also been observed in microgravity experiments performed on the ISS in both RF \citep{15eaa} and DC plasmas \citep{15jjj} with subsonic ion flow.

Future work, such as the inclusion of a variable electron density within the simulation region will refine these results and further enhance the use of dust grains to probe plasma conditions and quantify the difficult-to-measure changes in plasma parameters causing structural transitions of the dust grains.

\section*{ACKNOWLEDGMENTS}
This work is supported by the National Science Foundation under Grants No. 1707215, 1740203 and NASA/JPL contract 1571701. Computer simulations were carried out on the Kodiak HPC cluster with support provided by the Baylor University High Performance and Research Computing Services.


\end{document}